\documentclass{article}

\usepackage{arxiv}

\usepackage[utf8]{inputenc} 
\usepackage[T1]{fontenc}    
\usepackage{booktabs}       
\usepackage{amsfonts}       
\usepackage{nicefrac}       
\usepackage{microtype}      
\usepackage{graphicx}
\usepackage{doi}
\usepackage[all]{xy}
\usepackage{setspace}
\usepackage{xcolor}
\usepackage{tikz}
\usepackage{mathtools} 
\usepackage{mathrsfs}  
\usepackage{amsbsy}
\usepackage{amsmath}
\usepackage{amssymb}

\usetikzlibrary{cd}

\title{Internal tensorial variables and a heat transport equation with inertial, thermal viscosity and vorticity terms}

\author{Liliana Restuccia\\
University of Messina, \\Department of Mathematical and Computer
Sciences, Physical Sciences and Earth Sciences,\\ 
Viale F. Stagno d'Alcontres, Salita Sperone 31, 98166 Messina, Italy \\
Corresponding author: lrestuccia@unime.it\\
\And David Jou\\
Departament de Fis\'{i}ca, Universitat Aut\`onoma
de Barcelona,  \\ 08193  Bellaterra, Catalonia, Spain \\ 
\And and Michal Pavelka \\
 Mathematical Institute, Faculty of Mathematics and Physics, Charles University,\\ 
 Sokolovsk\'{a} 83, 18675 Prague, Czech Republic
}

\newcommand{\shorttitle}{Heat transport with vorticity}

\begin{document}

\def\defi{\stackrel{\hbox{\tiny \rm def}}{=}}
\newcommand{\di}{\displaystyle}
\newcommand{\pn}{\par\noindent}
\def\eq#1{(\ref{#1})}
\def\BE{\mathbf E}
\def\BPM{\mathbf P}
\def\BD{\mathbf D}
\def\BH{\mathbf H}
\def\BB{\mathbf B}
\def\BN{\mathbf n}
\def\BV{\mathbf v}
\def\BF{\mathbf F}
\def\JQ{\mbox{\boldmath $J$}}
\def\BET{\mbox{\boldmath $\xi$}}
\def\LAM{\Lambda}
\def\XII{\xi}
\def\MI{m^{(1)}}
\def\M2{m^{(2)}}

\maketitle

\begin{abstract}
Phonon hydrodynamics  describes the motions of heat carriers (phonons) at sub-continuum scales: diffusive, ballistic, viscous, and vortical. In a previous paper, these behaviours   were investigated  within the framework of non-equilibrium thermodynamics with internal variables at the macroscopic scale, deriving  generalizations of the Guyer-Krumhansl equation.  In particular, a generalized heat conduction equation, containing    not only    the Fourier, Maxwell-Vernotte-Cattaneo,  and  Guyer-Krumhansl contributions, but also  a term  describing phonon vortices, was obtained.  
In this paper, we provide new insight and clarifications into the same model for rigid heat-conducting media.   Then,  we  a posteriori identify 
two  non-local macroscopic internal variables, $\mathbf{Q}^{s}$   and  $\mathbf{Q}^a$ (the symmetric part  and  the antisymmetric part of a second order tensor $\mathbf{Q}$) with 
 the symmetric (changed in sign) and antisymmetric gradients of the heat flux, $-(\nabla\mathbf{J}^{(q)})^{s}$ and $(\nabla\mathbf{J}^{(q)})^{(a)}$.
 Also an   identification  of these two tensorial internal variables is obtained by an asymptotic approach.  
This generalizes the heat equation with additional terms  containing the time derivative of the heat flux  describing the viscous and vortical motions of phonons.   These terms may describe the transfer from ordered rotational motion of phonon vortices to rotational microscopic motions of diatomic particles constituting complex polar crystals, in analogy to the hydrodynamics of classical micropolar fluids. Therefore, this paper fits into the currently explored area of phonon vorticity and its interactions with the heat flux itself. 
\end{abstract}

\tableofcontents

\doublespace

\section{Introduction}
The recent interest in phonon hydrodynamics \cite{a}-\cite{c} and, in particular, in heat vortices  
\cite{21}-\cite{24} has fostered several extensions of the Guyer-Krumhansl thermal transport equation   
\cite{a}-\cite{c}, \cite{12}, \cite{14}, \cite{15}.
 In particular, in \cite{1} a generalized Guyer-Krumhansl thermal transport equation for the heat flux in heat rigid conductors was obtained in the framework of non-equilibrium thermodynamics with internal variables (NET-IV) (see \cite{Pr}-\cite{d}), by using nonlocal vectorial variables at sub-continuum scale, and accounting for vortical heat effects. In usual phonon hydrodynamics, the shear viscosity of phonon fluid is taken into account \cite{a}-\cite{c}. However, the presence of phonon vortices suggests the convenience of enlarging such hydrodynamic framework by taking into account some rotational viscosity of the phonon fluid, analogous to that arising in micropolar fluids  \cite{e}, \cite{f}, \cite{g}. 
 \par The existence of vortices does not imply the need of a rotational viscosity; vortices arise in the usual Navier-Stokes equation with only shear viscosity (or with shear viscosity and bulk viscosity). The essential physical aspect of rotational viscosity is the irreversible transfer of some part of the macroscopic or mesoscopic rotational energy of vortices to microscopic rotational energy of the constituent particles \cite{e}, \cite{f}, \cite{g}. In the case of micropolar fluids, this refers to elongated molecules which may rotate with respect to some of their axes, and which take organized macroscopic rotational energy into disorganized molecular rotations. In the case of phonon hydrodynamics, such microscopic rotations could arise in the case that the basic constituents of the crystal are not single particles, but couples of particles which do not only vibrate around their respective equilibrium positions but also rotate around their equilibrium orientation. In such a case, a part of the organized rotational momentum of the phonon fluid would go into disorganized rotational motion of the couples constituting the crystal. Thus, it would be found not in any crystal, but in particular crystals having such microscopic constitutions.
\par In this sense, we are enlarging the concept of phonon hydrodynamics to more complex and subtle interpretations than the simplest phonon hydrodynamics. Some attempts in this direction have been done by analogy of some rheological models describing non-Newtonian fluids, as power-law fluids \cite{h} or non-linear viscoelastic fluids \cite{l}, but considering the symmetric part   $\mathbf{Q}^{s}$  of  an  tensorial internal variable $ \mathbf Q$. Here, instead, we also consider the  antisymmetric part  $\mathbf{Q}^a$ of such a tensor and  we 
identify   $\mathbf{Q}^{s}$   and  $\mathbf{Q}^a$,    with 
 $-(\nabla\mathbf{J}^{(q)})^{s}$ and $(\nabla\mathbf{J}^{(q)})^{(a)}$, 
the symmetric (changed in sign) and  antisymmetric parts of the gradient of the heat flux, respectively.  Also an   identification  of these two tensorial internal variables is obtained by an asymptotic approach.
\par The presence of non-local effects in generalized equations for heat transport is of special interest in systems in which  $l$  (the mean-free path of the heat carriers, phonons) is comparable to (or bigger than) $L$, the  size of the system (for instance, it could be the radius of a cylindrical channel), 
i.e., for Knudsen number  $\frac{l}{L}\geq1$. This may be valid  for    
 systems  small enough,  as the  nanosystems,  having  their size   comparable to the mean free path, or 
when (see (\cite{1})) "the  mean free path is long enough to become  comparable to the size of the system, 
 as for  low-temperature phonons (superfluid helium-4 \cite{landau41}),  or in rarefied  systems, as for instance  it occurs  in aerodynamics   at low environmental pressure". 
\par In   nanosystems, generalizations of the  Guyer-Krumhansl equation (see \cite{GK1},  \cite{GK2}  and  \cite{6}) for the heat transfer are used.
 In  \cite{7}, a  two-fluid ballistic-diffusive hydro-dynamic model  is formulated at the nanoscale (see also  \cite{landau41}, \cite{Che01a}, \cite{Che02a}.
In \cite{8}, an  approach is  developed for heat rigid conductors, introducing a single vectorial internal variable and   postulating a generalized form for the entropy current.
In   \cite{9}, in the case of one-dimensional isotropic rigid heat conductors, 
 a generalized ballistic-conductive heat transport equation  is derived.  In     \cite{10}  the three-dimensional case  of the model studied in  \cite{9}
was treated. 
In  \cite{1},  a generalized 
Guyer-Krumhansl  equation was worked out, using  two non local vectorial (or tensorial)  internal variables.  For some remarks about the internal variables and some versions of non-equilibrium thermodynamics see also Section 2 of Reference \cite{1}.
 In  \cite{24} a description of thermal vortices of phonons was given by the GENERIC (General
Equation for Non-Equilibrium Reversible-Irreversible Coupling) approach (see \cite{go,og,hco,PKG}). 
\par In this paper  we have used the    Gyarmati approach  (see \cite{G1}),  where  the specific entropy $s$   depends on the  equilibrium  state variables  but also presents   homogeneous quadratic  terms depending  on vectorial, tensorial variables, and current densities,  to  model  irreversible processes.  We have taken into account the model developed in  \cite{1}    and we have   identified  from the physical point of view the  two   internal variables  introduced   in \cite{1}  in the thermodynamic state space,  
\par The paper is organized as follows. In Section 2, we  recall,    with  further clarifications and insights,   the  model for  rigid media utilized in \cite{1},  
  introducing    in the thermodynamic state space  not only  the  internal energy   and    the heat flux $\mathbf{J}^{(q)}$, 
but also two  non-local macroscopic internal variables, $\mathbf{Q}^{s}$   and  $\mathbf{Q}^a$,  the symmetric part  and  the antisymmetric part of a second order tensor $\mathbf{Q}$.  
Using the usual formalism of Gyarmati approach, the evolution equations of  $\mathbf{Q}^{s}$ and   
$\mathbf{Q}^{(a)}$,  with  $\mathbf{Q}^{(a)}$ the related axial vector to $\mathbf{Q}^{a}$, are derived. When neglecting some terms in them, $\mathbf{Q}^{s}$ and 
$\mathbf{Q}^{(a)}$ become approximately equal to  the symmetric and antisymmetric parts of the gradient of the heat flux   $\mathbf{J}^{(q)}$, which leads to an evolution equation for the heat flux  as in  \cite{1}.  This equation represents a generalized Guyer-Krumhansl equation with kinematic and viscous motions of phonons where also a rotational viscosity is present. 
In Section 3, other different generalized heat equations are derived in the  same  physical  situation as in  \cite{1},   assuming that the tensorial quantities $\mathbf{Q}^s$ and $\mathbf{Q}^{(a)}$ relax quickly to their respective quasi-equilibria, characterized by
$\dot{\mathbf{Q}}^s \approx 0 \quad\mbox{and}\quad \dot{\mathbf{Q}}^{(a)} \approx 0.$
 In Section 4, using the evolution equations for  $\mathbf{Q}^s$ and $\mathbf{Q}^{(a)}$, we identify the two internal tensorial variables  $\mathbf{Q}^s$ and $\mathbf{Q}^{(a)}$   with 
 $-(\nabla\mathbf{J}^{(q)})^{s}$ and $(\nabla\mathbf{J}^{(q)})^{(a)}$, respectively, and we explore the subsequent evolution equation for the heat flux, obtaining a heat transfer  equation  where there  are additional terms  containing the time derivative of the heat flux (viscous and vortical motions of the phonons). This situation is possible in complex solid media where the molecules have rotational motions around their equilibrium position that interact with the vortical motion of phonons.  Finally, Section \ref{sec.asymp} contains an alternative method for the identification of the internal tensorial fields $\mathbf{Q}^{s}$ and $\mathbf{Q}^{a}$,     based on  an asymptotic approach  consisting in  approximately solving the evolution equations  for the internal variables  while keeping the zero-th and first orders.

\section{A  thermodynamic formulation with internal variables, describing non-local and vortical thermal motions  of phonons }
 In this paper, we take into account  the same model for a rigid heat-conducting media used in \cite{1}, that   here we recall  with new insights and   clarifications, 
and we  introduce   in the thermodynamic state space $C$,  not only  the  internal energy   and    the heat flux $\mathbf{J}^{(q)}$, but also two  non-local macroscopic internal variables, $\mathbf{Q}^{s}$   and  $\mathbf{Q}^a$,  the symmetric part  and  the antisymmetric part of a second order tensor $\mathbf{Q}$, i.e.:
$
 C = C\left( u,\mathbf{J}^{(q)}, \mathbf{Q}^{s},  \mathbf{Q}^{a}\right).
$
   \par Thus, as in   \cite{1}, we consider  the following balance equations:  
   \begin{itemize}
 \item the \textit{mass conservation law, }    having the form
\begin{equation}\label{rho}
\dfrac{d \rho}{d t}\,+  \rho\,\nabla\cdot{\mathbf v } \,=\,0, 
\end{equation}
 where  $\rho$  is  the mass density field   of the  considered medium,   
  $d/dt$ is  the  material  time derivative,   defined  by
$\;\frac{d }{d t}\,=\,\frac{\partial}{\partial t}\,+
\,v_i\,\frac{\partial }{\partial x_i}\,$    with  $\frac{\partial}{\partial t}$  and $\frac{\partial }{\partial x_i}$ the partial temporal derivative and the  partial spatial derivative,   $v_i$ is the velocity field and the symbol  
"$\nabla\cdot$"   and   "$\nabla$"  denote the divergence  operator and the gradient operator;
 \item the \textit{internal energy balance equation, }  written in the  form:
\begin{equation}\label{u}
\rho\,\dfrac{d u}{d t}\,+\,\nabla\cdot\mathbf{J^{(q)}}\,=\,0\,,
\end{equation}
where $u$  is the internal energy density,  $\mathbf{J^{(q)}}$ is the heat flux vector, the current density of the internal energy  and the  heat source is disregarded;\\
   \item the \textit{entropy inequality,}  given  by 
\begin{equation}
\label{A}
 \sigma ^ {(s)}=\rho\,\dfrac{d s}{d t}\,+\,\nabla\cdot  \mathbf{J^{(s)}} \geqslant0, 
\end{equation}
where the fields of specific entropy $s$, the  entropy flux density $\mathbf{J}^{(s)}$ and the  entropy production density $\sigma^ {(s)}$  are constitutive quantities,  functions of the  independent variables of the  thermodynamics state space  (see for comments for instance \cite {Mu1} and \cite {Mu2}).
\end{itemize}
\par In \cite{1}
a  Gyarmati approach  was used, where  
the specific entropy $s$ is not function only on the local equilibrium state variables, but also  depends in an homogeneous quadratic way on  $\mathbf{J}^{(q)}$,   $\mathbf{Q}^{s}$ and  $\mathbf{Q}^a$ in the following way 
\begin{equation}\label{ss2}
s\left( u,\mathbf{J}^{(q)}, \mathbf{Q}^{s},  \mathbf{Q}^{a}\right) = s^{(eq)}(u)- \frac{1}{2}\alpha_1 \left( \mathbf{J}^{(q)}\right) ^2 -\frac{1}{2}\alpha_2^{s}\mathbf{Q}^{s}:\mathbf{Q}^{s}  -\frac{1}{2}\alpha_2^{a}\mathbf{Q}^{a}:\mathbf{Q}^{a},
\end{equation}
where  $s^{(eq)}(u)$ is the equilibrium entropy function,  $\alpha_1$,  $\alpha_2^{s}$ and $\alpha_2^{a}$  are inductivity constants,  required positive   for the thermodynamic stability, and the symbol ":"  denotes double contraction, i. e.  for example  $ \mathbf{Q}^{s}:\mathbf{Q}^{s} = Q^{s}_{ij}Q^{s}_{ij}$.
\par Defining 
 \begin{equation} \label{def}
\left\{
\begin{aligned}
\dfrac{1}{T}\,&=\,\dfrac{\partial s}{\partial u}\\
\dfrac{\partial s}{\partial J^{(q)}_i}\,&=\,-\alpha_1 J^{(q)}_i, \\
\dfrac{\partial s}{\partial  Q_{ij}^{s}}\,&=\,-\alpha_2^{s} Q_{ij}^{s}, \\
\dfrac{\partial s}{\partial   Q_{ij}^{a}}\,&=\,-\alpha_2^{a} Q_{ij}^{a} , \\
\end{aligned} \right.
\end{equation}
as in  \cite{1},   we  have  assumed   the  entropy flux   $\mathbf{J}^{(s)}$     having  the form
\begin{equation}\label{E8}
\mathbf{J}^{(s)} =  T^{-1}\mathbf{J}^{(q)} - \beta_1^{s}\mathbf{Q}^{s}\cdot\mathbf{J}^{(q)}
- \beta_1^{a} \mathbf{Q}^{a}\cdot\mathbf{J}^{(q)} =  T^{-1}\mathbf{J}^{(q)} - \beta_1^{s}\mathbf{Q}^{s}\cdot\mathbf{J}^{(q)}
- \beta_1^{a} \mathbf{Q}^{(a)}\times \mathbf{J}^{(q)},  
 \end{equation}
where   $\beta_1^{s}$ and $\beta_2^{s}$ are constant  material coefficients, that conveniently represent the deviation of $\mathbf{J}^{(s)}$ from
the local equilibrium,  the symbol  "$\cdot$"  indicates   one  contraction, i. e. for example 
 $ \mathbf{Q}^{s}\cdot\mathbf{J}^{(q)}$   in components has the form  $Q_{ij}^{s} J_j^{(q)}, $ and  we  have  used 
the formula   $\mathbf{Q}^{a}\cdot \mathbf{J}^{(q)} =   \mathbf{Q}^{(a)}\times \mathbf{J}^{(q)}$, with  the symbol " $\times$ "  indicating the  vector product,  because it is  possible to associate  to every antisymmetric tensor, in our case $  \mathbf{Q}^{a},$  one and only one axial vector   $\mathbf{Q}^{(a)}$  (see  Appendix A).
How could tensors $\mathbf{Q}^s$ and $\mathbf{Q}^{a}$ be understood from the microscopic point of view? In the kinetic theory of phonons \cite{M1}  the second moment of the phonon distribution function can be considered as a state variable and can be represented by the symmetric tensor $\mathbf{Q}^s$. For the skew-symmetric tensor field, we have to go beyond the one-particle kinetic theory, moments of which are always symmetric in their indeces. The two-particle kinetic theory \cite{PKG} brings the tensor field of nonlocal vorticity (the second-order mixed moment of the distribution function with respect to the spatial distance of two particles and their relative velocity). The nonlocal vorticity tensor could be interpreted as related to the skew-symmetric field $\mathbf{Q}^a$.
\par In   \cite{1}
 the evolution equations for the independent variables $\mathbf{J}^{(q)}$,  $\mathbf{Q}^{s}$ and $\mathbf{Q}^{(a)}$   were found, using   the  definitions   (\ref{def}),    calculating  the entropy differential, 
 the time derivative of the entropy  in  the form  
\begin{equation}\label{E1AA}
\dot{s} =  T^{-1}  \dot{u}  -  \alpha_1\mathbf{J}^{(q)}\cdot \dot{\mathbf{J}}^{(q)}   - \alpha_2^{s}\mathbf{Q}^{s}:  \dot{\mathbf{Q}}^{s} -\alpha_2^{a}\mathbf{Q}^{a}:   \dot{\mathbf{Q}}^{a}, 
  \end{equation}
  (where the upper dot  "$^.$"   denotes the total time derivative)
  and  
 the  entropy production   $\sigma^{s}$ in the form  (see  (\ref{A})) 

 \begin{align}\label{sig} 
     \sigma^s =&  -  \rho\alpha_1\mathbf{J}^{(q)}\cdot\dot{\mathbf{J}}^{(q)}   +   \mathbf{J}^{(q)} \cdot \nabla T^{(-1)} - \beta_1^{s} \left[ \mathbf{J}^{(q)}\cdot\left( \nabla\cdot\mathbf{Q}^{s} \right)    
+ \mathbf{Q}^{s}: \left( \nabla\mathbf{J}^{(q)}\right)^{s}\right]  \nonumber\\
     &- \beta_1^{a}\nabla\cdot\left( \mathbf{Q}^{(a)}\times \mathbf{J}^{(q)}\right)   - \rho\alpha_2^{s}\mathbf{Q}^{s}:  \dot{\mathbf{Q}}^{s} -\rho\alpha_2^{a}\mathbf{Q}^{a}:\left(\dot{\mathbf{Q}}^{a}\right). 
       \end{align}
 In   (\ref{sig})    the   formula 
\begin{equation}\label{comp1}
\nabla\cdot\left( \mathbf{Q}\cdot \mathbf{J}^{(q)}\right)  = \mathbf{J}^{(q)}\cdot\left(\nabla\cdot \mathbf{Q}^T\right)  +  \mathbf{Q} : \left( \nabla\mathbf{J}^{(q)}\right)^T,
\end{equation}    
is used,  
valid for any tensor $\mathbf{Q}$  of second order (see (\ref{nablacdotA})   of Appendix B),  being $\mathbf{Q}^T $  the transposed tensor of  $\mathbf{Q} $  (in components   
$(Q_{ji})^T= Q_{ij}$).
\par Thus, in   (\ref{sig})  the formula  (\ref{comp1}) is applied  in the case of  $\mathbf{Q}^{s}$,  furthermore,  
$\nabla\mathbf{J}^{(q)}$  is decomposed  in its   symmetric    part   $\left( \nabla\mathbf{J}^{(q)}\right)^{s}$  and its  antisymmetric part $\left( \nabla\mathbf{J}^{(q)}\right)^{a}.$  
\par In components  $(\nabla\mathbf{J}^{(q)})^{s}$   and  $(\nabla\mathbf{J}^{(q)})^{a}   $   have    the following expressions: 
\begin{equation}\label{comp}
\small \left[  \left( \nabla\mathbf{J}^{(q)}\right)^{s}\right] _{ij} = \frac{1}{2}\left[  \left( \mathbf{J}^{(q)}\right)_{i,j} +  \left( \mathbf{J}^{(q)}\right) _{j,i}\right], \quad \left[  \left( \nabla\mathbf{J}^{(q)}\right) ^{a}\right] _{ij} = \frac{1}{2}\left[ \left( \mathbf{J}^{(q)}\right)_{i,j} - \left( \mathbf{J}^{(q)}\right) _{j,i}\right],  i,j =1,2,3,
\end{equation}
 where  a comma in lower indices denotes  the  partial spatial derivate for instance $ (\mathbf{J}^{(q)})_{i,j} = \frac{\partial J^{(q)}_{i}}{\partial x_j}=\left( \nabla\mathbf{J}^{(q)}\right)_{ij }. $
\par  In    \cite{1}  we  have  considered the case of a rigid heat conductor at rest, so that the substantial  time derivative is equal to the partial temporal derivative $\frac{\partial}{\partial t}$. 
Then, by virtue of the expression 
\begin{equation}\label{exp}
    \nabla\cdot\left( \mathbf{Q}^{(a)}\times \mathbf{J}^{(q)}\right)  = \mathbf{J}^{(q)}\cdot\nabla \times \mathbf{Q}^{(a)}  -  \mathbf{Q}^{(a)}\cdot \nabla\times \mathbf{J}^{(q)} 
\end{equation} (see Appendix D)
in  \cite{1} the following expression  for $ \sigma^s$ was found
 \begin{align}\label{E3}
     \sigma^s =& \mathbf{J}^{(q)} \cdot\left[ \nabla T^{(-1)} - \rho\alpha_1\dot{\mathbf{J}}^{(q)} - \beta_1^{s}\nabla
\cdot\mathbf{Q}^{s} - \beta_1^{(a)} \nabla
     \times\mathbf{Q}^{(a)}\right]    + \mathbf{Q}^{s}:\left[ -\rho\alpha_2^{s} \dot {\mathbf{Q}}^{s}  - \beta_1^{s}\left( \nabla\mathbf{J}^{(q)}\right)^{s}\right]  + \nonumber\\
     &+\mathbf{Q}^{(a)}\cdot\left[  -\rho\alpha_2^{(a)} \dot{\mathbf{Q}}^{(a)}  + \beta_1^{(a)}\left( \nabla\times \mathbf{J}^{(q)}\right) \right],
\end{align}
where  the symbol "$^.$"  indicates the derivative with respect to time. 
\par  We obtain  the identifications 
 $\beta_1^a = \beta_1^{(a)}$, and $\alpha_2^{(a)}=2\alpha_2^a$, 
being the skew-symmetric tensor $\mathbf{Q}^a$  represented by the axial vector $\mathbf{Q}^{(a)}$, related to that tensor via  the  formula (see (\ref{Qab}) of Appendix A)
\begin{equation}\label{Qaij}
Q^a_{ij} = -\epsilon_{ijk} Q^{(a)}_k, 
\end{equation}
and 
\begin{equation}\label{E3a}Q^a_{ij}\dot{Q}^a_{ij} = \epsilon_{ijk}Q^{(a)}_k \epsilon_{ijl}\dot{Q}^{(a)}_l = (\delta_{jj}\delta_{kl}-\delta_{lj}\delta_{kj})Q^{(a)}_k\dot{Q}^{(a)}_l = 3 \mathbf{Q}^{(a)}\cdot\dot{\mathbf{Q}}^{(a)} - \mathbf{Q}^{(a)}\cdot\dot{\mathbf{Q}}^{(a)} = 2\mathbf{Q}^{(a)}\cdot\dot{\mathbf{Q}}^{(a)}.  
\end{equation}
\par 
The entropy production  $\sigma^s$ (\ref{E3}) is given by the sum of products of thermodynamic fluxes and forces.  
Assuming  that  the media under consideration are  perfectly isotropic and thus have the symmetry properties of being invariant under orthogonal transformations, 
 we obtain the  following    linear relations  among  the  chosen thermodynamic  fluxes: 
  $\mathbf{J}^{(q)}$, $\mathbf{Q}^{s}$, $\mathbf{Q}^{(a)}$  and the  corresponding  thermodynamic affinities (see \cite{1}):
\begin{subequations}\label{eq.evo}
 \begin{align}\label{E3B}
     \mathbf{J}^{(q)} =&\lambda_1T^2\left[ \nabla T^{(-1)} - \rho\alpha_1\dot{\mathbf{J}}^{(q)} - \beta_1 ^{s}\nabla
\cdot\mathbf{Q}^{s} - \beta_1^{(a)}\nabla
\times\mathbf{Q}^{(a)}\right] ,   \\ 
\label{E3C}
     \mathbf{Q}^{s} =& \lambda_2^{s}\left[ -\rho\alpha_2 ^{s}\dot{\mathbf{Q}}^{s}  - \beta_1^{s}(\nabla\mathbf{J}^{(q)})^{s}\right] ,   \\
 \label{E3D}
\mathbf{Q}^{(a)} =&  \lambda_2^{(a)}\left[  -\rho\alpha_2^{(a)} \dot{\mathbf{Q}}^{(a)}  + \beta_1 ^{(a)}(\nabla\times (\mathbf{J}^{(q)})\right] ,
\end{align}
\end{subequations}
where  $\lambda_1, \lambda_2^{s}, \lambda_2^{(a)}$  are  positive scalar coefficients,  with  $\lambda_1$ the heat conductivity  coefficient,  and  $\mathbf{J}^{(q)}$,   $\mathbf{Q}^{s}$    and  $\mathbf{Q}^{(a)}$   are  a  polar vector,    a second order tensor and    an axial vector,   respectively,  that  do not  couple for the Curie principle. 
\par The two equations  (\ref{E3C}) and   (\ref{E3D}) are not independent equations but are part of the system of equations (\ref{E3B})-(\ref{E3D}). 
 Equation (\ref{E3B}) relates the temporal derivative of the heat flux with the divergence of $\mathbf{Q}^s$. This  could  suggest that $\mathbf{Q}^s$ may be interpreted as the symmetric part of the flux of the heat flux, or some quantity proportional to it, as it is a priori assumed in some formulations of nonlocal extended thermodynamics
  \cite{12},   \cite{M2}. Then, $\mathbf{Q}^a$ could be analogously interpreted as being proportional  to the antisymmetric part of the flux of the heat flux, which we have introduced here for the first time.
\par Taking into consideration that  (see expression   $(\ref{comp})_1$)
$$\nabla\cdot\left[  \left( \nabla\mathbf{J}^{(q)}\right)^{s}\right]  =  \frac{1}{2}  \nabla \left( \nabla\cdot \mathbf{J}^{(q)}\right)   +  \frac{1}{2}\Delta \mathbf{J}^{(q)},$$
 neglecting  in  the evolution equations (\ref{E3C}) and (\ref{E3D})  the terms in  
$\dot{\mathbf{Q}}^{s}$ and  $\dot{\mathbf{Q}}^{(a)}$, respectively, 
inserting the obtained approximate  evolution equations for  $\mathbf{Q}^{s}$ and  $\mathbf{Q}^{(a)}$ in the expression  ( \ref{E3B}),  in \cite{1} 
it was  derived the    following final result concerning a generalized  Guyer-Krumhansl heat equation 
\begin{equation}\label{E102L} 
\mathbf{J}^{(q)} =  - \lambda_1  \nabla T   -   \tau \dot {\mathbf{J}}^{(q)} + l_1^2\left[ \nabla \left( \nabla\cdot \mathbf{J}^{(q)}\right) +\Delta\mathbf{J}^{(q)} \right]  
- l_2^2\nabla\times(\nabla\times\mathbf{J}^{(q)}), 
\end{equation}
with  \begin{equation}\label{cost2}
\tau =\rho\lambda_1 T^2\alpha_1,\quad l_1^2 = \frac{1}{2}\lambda_1 T^2 \lambda_2^{s} (\beta_1^{s})^2,\quad 
l_2^2 = \lambda_1 T^2\lambda_2^{(a)} (\beta_1^{(a)})^2, 
\end{equation} 
Note that the smallness of $\mathbf{Q}^{s}$ and $\mathbf{Q}^{(a)}$ should be understood in the sense of the smallness of the non-dimensional quantities identified later in Section \ref{sec.asymp}. 
Equation (\ref{E102L}) may be of special interest for the researchers on phonon-hydrodynamics.  Indeed, it is the usual Guyer-Krumhansl equation, with in  the third term on the right-hand side being related to the viscous behaviour of phonon motions, and the fourth term on the right-hand side is new, and it describes the rotational viscous behaviour which arises in rotational motions of the phonon fluid, namely, in  heat vortices. This term may be physically relevant when the particles constituting the crystal are not single particles but dipoles (see Fig. 1). In this case, the rotational viscosity describes the transfer from organized rotational motion of the heat vortex to disorganized individual rotations of the dipoles over their respective axes.
\begin{figure}[h!]
\centerline{\includegraphics[width=0.4\textwidth]{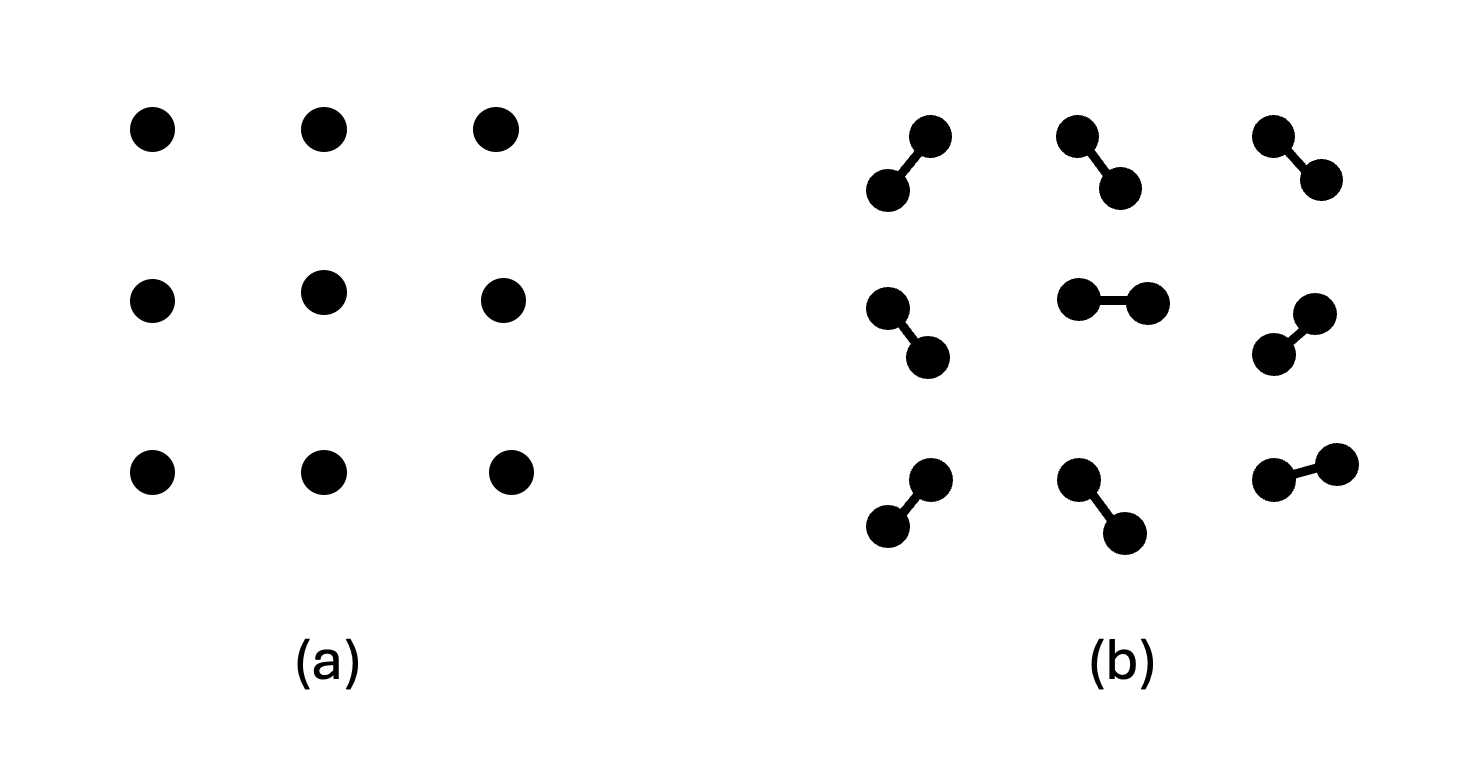}}
\caption{" The dots stand for particles. In (a), the crystal is made by single particles; in (b), it is made by dipoles.  In (a), the particles vibrate; in (b), they vibrate and rotate".}
\label{fig1}
\end{figure}
\par Equation (\ref{E102L}) can be written in the form 
 \begin{equation}\label{E102LL} 
\dot{\mathbf{J}}^{(q)}      =    -   \frac{\mathbf{J}^{(q)}}{\tau}   - \frac{\lambda_1 }{\tau} \nabla T  + \nu_1\left[ \Delta\mathbf{J}^{(q)} +\nabla \left( \nabla\cdot \mathbf{J}^{(q)}\right) \right]
- \nu_2\nabla\times(\nabla\times\mathbf{J}^{(q)}), 
\end{equation}
 with  $\nu_1$ and $\nu_2$ playing the role of shear and kinematic viscosities, given by 
 $\nu_1=  \frac{l_1^2 }{\tau}$  and  $\nu_2= \frac{l_2^2 }{\tau}.$  In the left hand side,  the velocity 
   $\mathbf{v}$ is the average velocity of phonon flow, given by  $\mathbf{v} = \frac{\mathbf{J}^{(q)}}{(\rho c_vT)}$, with $c_v$ the specific heat per unit mass (see for instance  \cite{24}).
\par  From the point of view of hydrodynamic analogies,   equation (\ref{E102LL}) 
 is  similar to the following  Navier-Stokes equation for micropolar fluids, having elongated, not spherical molecules,  with rotational  degrees of freedom, that  could vibrate with respect to their equilibrium position, giving contributions to the rotational energy, interacting with phonon vortices  (see \cite{56},  \cite{57}, \cite{58}, \cite{59})
 \begin{equation}\label{E102LLL} 
\dot{\mathbf{v}}  =    -   \frac{\nabla p}{\rho}   + \frac{\mu_1}{\rho} \Delta\mathbf{v}  + \frac{\mu_0}{\rho} \nabla \left(\nabla\cdot \mathbf{v}\right)  
+ \frac{\mu_2}{\rho}\left[ \nabla\times(\nabla\times\mathbf{v}) - 2\bf{\omega}\right], 
\end{equation}
 where $\mathbf{v}$ is the velocity field, $\nabla p$ the pressure field, $\mu_0,\mu_1$ and  $\mu_2$  are the bulk, 
 the shear and the rotational viscosities and $\bf{\omega}$ is the local (non quantum) spin or intrinsic angular momentum, which we do not have considered in this paper,  of the micropolar fluid.   In  (\ref{E102LLL}) $\bf{\omega}$  appears as related to the balance equation of the  angular momentum. In   (\ref{E102LL})   we take $\bf{\omega} = 0$.
If  the liquid flows in a porous environment,   an   additional term $-\frac{\alpha\mathbf{v}}{\rho}$ in the right hand side of Navier-Stokes equation  (\ref{E102LLL})  could be present (a Darcy term), being $\alpha$  the porosity coefficient.  Such Darcy term would play a role analogous to the first term on the right  side-hand side of equation (\ref{E102LL}), namely, the term   $\frac{\mathbf{J}^{(q)}}{\tau}$. 
\par In  (\ref{E102L}) 
 the coefficients  $l_1^2$  and $l_2^2 $  are the square of mean free paths of phonons,  respectively,  related to the exchange of linear momentum and to internal angular momentum, $l_1^2$ is   multiplied by a term describing non-local effects and, in particular, the viscous motions of phonons. Coefficient $l_2^2$ is multiplied by a non-local term related to the rotational motion of the heat flux,  describing rotational viscous motion of the heat carriers in crystals of elongated or polarized constituents. These terms are analogous to the terms that in the Navier-Stokes equation for viscous fluids describe the shear and rotational viscous effects of fluid particles, as recalled in equation \eqref{E102LL}.
\par Equation (\ref{E102L}), in the case where    the coefficient  $l_2^2 $  can be neglected, reduces to Guyer-Krumhansl  heat equation
 \begin{equation}\label{E10G} 
\mathbf{J}^{(q)} =  - \lambda_1  \nabla T   -   \tau \dot {\mathbf{J}}^{(q)} + 
l_1^2 \left[ \nabla \left( \nabla\cdot \mathbf{J}^{(q)}\right) +\Delta\mathbf{J}^{(q)} \right],   
\end{equation}
describing at sub-continuum scale the ballistic motion of phonons, that are subjected to collisions with the walls of the solid. 
 \par Equation (\ref{E102L}), in the case in which the coefficients  $l_1^2$ and    $l_2^2 $  
 can be disregarded,  reduces to the Maxwell-Vernotte-Cattaneo equation 
  \begin{equation}\label{E10M} 
  \tau \dot {\mathbf{J}}^{(q)} + \mathbf{J}^{(q)} =  - \lambda_1  \nabla T, 
\end{equation}
describing  at macroscopic scale  thermal signals (second-sound phenomena) having a relaxation time  $\tau$ and a finite velocity of propagation and   at sub-continuum scale  the diffusive motion of phonons colliding with the core of the solid;  in the case  
 where the coefficients $\tau,$  $l_1^2  $  $l_2^2 $ 
 can be neglected   we obtain  the Fourier  equation,
describing thermal signals with   relaxation time $\tau$  null, valid at large space scales and at low frequencies. 

\section{Various levels of beyond-Fourier heat conduction  }
Let us show how Equation (\ref{E102L}) is connected with various levels of description of beyond-Fourier heat conduction.
\par  First, Equations   (\ref{E3B}), (\ref{E3C}) and (\ref{E3D}) can be written also in the following form
\begin{subequations}
 \begin{equation} \label{E3Ba}
\tau_1\dot{\mathbf{J}}^{(q)} + \mathbf{J}^{(q)} =\lambda_1T^2\left[ \nabla T^{(-1)}  - \beta_1 ^{s}\nabla
\cdot\mathbf{Q}^{s} - \beta_1^{(a)}\nabla
\times\mathbf{Q}^{(a)}\right] ,   
\end{equation}
\begin{equation}\label{E3Bb}
\tau_2\dot{\mathbf{Q}}^{s} =  - \mathbf{Q}^{s} - \lambda_2^{s} \beta_1^{s}(\nabla\mathbf{J}^{(q)})^{s},   
\end{equation}
\begin{equation}\label{E3Bc}
  \tau_3 \dot{\mathbf{Q}}^{(a)}=      -     \mathbf{Q}^{(a)} + \lambda_2^{(a)} \beta_1 ^{(a)}(\nabla\times \mathbf{J}^{(q)}), 
\end{equation}
\end{subequations}
  with 
 \begin{equation} \label{tau}
 \tau_1  =\tau = \rho\lambda_1 \alpha_1 T^2 \quad  \text{(see}  (\ref{cost2})_1), \quad \tau_2=\rho \lambda_2^{s}\alpha_2 ^{s},  \quad\tau_3 =\rho\lambda_2^{(a)}\alpha_2^{(a)}
 \end{equation}
   the relaxation times of  $\mathbf{J}^{(q)}$,  $\mathbf{Q}^{s}$ and $\mathbf{Q}^{(a)},$ respectively. These equations represent the most detailed level of description where three state variables and three evolution equations are present (on top of the energy conservation equation).

\par Detailed evolution equations (\ref{E3Bb}) and  (\ref{E3Bc}) can be then reduced by assuming that the tensorial quantities $\mathbf{Q}^s$ and $\mathbf{Q}^{(a)}$ relax quickly to their respective quasi-equilibria, characterized by
\begin{equation}
\tau_1\approx 0,\quad \tau_2\approx 0,\quad \text{then }\quad 
    \tau_1\dot{\mathbf{Q}}^s \approx 0 \quad\mbox{and}\quad \tau_2\dot{\mathbf{Q}}^{(a)} \approx 0.
\end{equation}
Again, the smallness of $\mathbf{Q}^{s}$ and $\mathbf{Q}^{(a)}$ should be understood in the sense of the smallness of the non-dimensional quantities identified later in Section \ref{sec.asymp}.
\par The evolution equations for $\mathbf{Q}^s$ and $\mathbf{Q}^{(a)}$ then become the constitutive relations
\begin{equation}\label{eq.Q0}
     \mathbf{Q}^{s} = -\lambda_2^{s}\beta_1^{s}(\nabla\mathbf{J}^{(q)})^{s}  
     \qquad\mbox{and}\qquad
\mathbf{Q}^{(a)} =  \lambda_2^{(a)}\beta_1 ^{(a)}\nabla\times \mathbf{J}^{(q)}.
\end{equation}
When we plug these constitutive relations into the remaining evolution equation for the heat flux (Equation \eqref{E3Ba}), we obtain 
\begin{align}\label{eq.reduced}
    \mathbf{J}^{(q)} =&\lambda_1T^2\left[ \nabla T^{(-1)} - \rho\alpha_1\dot{\mathbf{J}}^{(q)} 
    + (\beta_1 ^{s})^2\lambda_2^{s}\nabla\cdot (\nabla\mathbf{J}^{(q)})^{s} 
    - (\beta_1^{(a)})^2\lambda_2^{(a)}\nabla \times\left(\nabla\times \mathbf{J}^{(q)}\right)\right]\nonumber\\
    =& \lambda_1T^2\left[ \nabla T^{(-1)} - \rho\alpha_1\dot{\mathbf{J}}^{(q)} 
    + \left(\frac{1}{2}(\beta_1 ^{s})^2\lambda_2^{s}-(\beta_1^{(a)})^2\lambda_2^{(a)}\right)\nabla(\nabla\cdot \mathbf{J}^{(q)}) \nonumber\right.\\
    &\left.\qquad\qquad +\left(\frac{1}{2}(\beta_1 ^{s})^2\lambda_2^{s}+(\beta_1^{(a)})^2\lambda_2^{(a)}\right)\Delta \mathbf{J}^{(q)}\right],
\end{align}
which is actually the generalized Guyer-Krumhansl equation \eqref{E102L}.
\par In case of $\beta_1^{(a)}=0$, when the entropy does not depend on the skew-symmetric tensor $\mathbf{Q}^a$, Equation \eqref{eq.reduced} simplifies to the Guyer-Krumhansl equation,
\begin{equation}
    \mathbf{J}^{(q)} =\lambda_1T^2\left[ \nabla T^{(-1)} - \rho\alpha_1\dot{\mathbf{J}}^{(q)} 
    + (\beta_1 ^{s})^2\lambda_2^{s}\nabla\cdot (\nabla\mathbf{J}^{(q)})^{s} \right].
\end{equation}
Moreover, if we assume that the entropy is independent of the symmetric tensor $\mathbf{Q}^s$ as well, the equation further simplifies to the Cattaneo equation \cite{catt},
\begin{equation}
    \mathbf{J}^{(q)} =\lambda_1T^2\left[ \nabla T^{(-1)} - \rho\alpha_1\dot{\mathbf{J}}^{(q)} \right].
\end{equation}
Finally, if the dependence of entropy on the heat flux is negligible as well, the equation restores the usual Fourier heat conduction,
\begin{equation}\label{F}
    \mathbf{J}^{(q)} = - \lambda_1\nabla T.
\end{equation}
In the following Section, we provide more detailed dynamics of $\mathbf{J}^{(q)}$ by an identification of the tensorial internal variables.

\section{A posteriori identification of the two internal variables  in a steady state} 
In this Section, we identify the two internal variables $\mathbf{Q}^s$ and   $\mathbf{Q}^{a}$ with the symmetric and antisymmetric parts of the gradient of the heat flux, respectively.
  \par Taking into consideration  (\ref{E3C}) and (\ref{E3D}),  we propose an a posteriori  interpretation of the physical meaning   of two  internal variables,
  \begin{equation}\label{E3DD}
  \mathbf{Q}^{s} =  - \left( \nabla\mathbf{J}^{(q)}\right) ^{s}, \quad \text{and}\quad
 \mathbf{Q}^{a} =    \left( \nabla\mathbf{J}^{(q)}\right) ^{a}.
 \end{equation}
 \par Then,   we consider that the entropy is a function $s(u, \mathbf{J}^{(q)},  -  \left(\nabla\mathbf{J}^{(q)}\right)^{s},  \left( \nabla\mathbf{J}^{(q)}\right) ^{a}$. 
 A comparison of formulations based on   $s(u, \mathbf{J}^{(q)}, \mathbf{Q}^s)$ and 
 on $ s(u,\mathbf{J}^{(q)}, \left(\nabla\mathbf{J}^{(q)}\right)^{s})$ was carried out in
 \cite{CF2007} and \cite{CJRV2014}. 
 Here,  we take into account also $\left( \nabla\mathbf{J}^{(q)}\right) ^{a}$,  furthermore  $-\left(\nabla\mathbf{J}^{(q)}\right)^{s}$  having  the minus sign,   and we follow a different approach, instead of searching independent evolution equations for $-\left(\nabla\mathbf{J}^{(q)}\right)^{s}$ and $\left( \nabla\mathbf{J}^{(q)}\right) ^{a}$.
Furthermore, if  we consider the related axial vector  $\mathbf{Q}^{(a)} =   \left(\nabla\mathbf{J}^{(q)}\right)^{(a)}$,  we obtain the following identification 
\begin{equation}\label{E3DDD}
\left( \nabla\mathbf{J}^{(q)}\right) ^{(a)}= \frac{1}{2}\left( \nabla\times\mathbf{J}^{(q)}\right) 
 \end{equation}
(see (\ref{comp}) and  Appendix C).
\par Thus, equations (\ref{E3Bb}) and (\ref{E3Bc}) take  following form
\begin{subequations}
\begin{equation}\label{E3Bbb}
\tau_2\dot{(\nabla\mathbf{J}^{(q)})}^{s} =  - (1- \lambda_2^{s}\beta_1^{s})(\nabla\mathbf{J}^{(q)})^{s},    
\end{equation}
\begin{equation}\label{E3Bcc}
\tau_3(\nabla\times \dot{\mathbf{J}}^{(q)}) =  -   \left( 1 - 2\lambda_2^{(a)} \beta_1 ^{(a)}\right) (\nabla\times \mathbf{J}^{(q)}). 
\end{equation}
\end{subequations}
Equations (\ref{E3Bbb}) and   (\ref{E3Bcc})  in the steady state  (where the terms in  $\tau_2\dot{(\nabla\mathbf{J}^{(q)})}^{s}$ and  $\tau_3(\nabla\times \dot{\mathbf{J}}^{(q)})$ are null)   are true   
when the multiplicative coefficients are  zero, i.e. 
$$1- \lambda_2^{s}\beta_1^{s}=0, \quad 1 -2\lambda_2^{(a)}\beta_1 ^{(a)}= 0,$$ being  
 $$\nabla\mathbf{J}^{(q)})^{s}\neq 0, \quad \nabla\times \mathbf{J}^{(q)}\neq 0, $$  the relaxation  times $\tau_2\neq 0$ and $\tau_3\neq 0, $  and  $\mathbf{J}^{(q)}$  is  a function dependent only on the spatial variables and not on the time variable t.
Therefore, the definitions a posteriori of internal variables are valid when
\begin{equation}\label{E3Bccc} \lambda_2^{s}\beta_1^{s}=1 \quad\mbox{and}\quad  \lambda_2^{(a)}\beta_1 ^{(a)}= 1/2.
\end{equation}
Although these conditions have been obtained in the steady state, they can be used in the general case as well because the coefficients are independent of the time-derivatives.
 \par  It follows that if we assume  the two internal variables as (\ref{E3DD})$_1$  and  (\ref{E3DDD}),  we obtain the conditions  (\ref{E3Bccc})$_1$   and (\ref{E3Bccc})$_2$ and 
 so we get that the definitions (\ref{E3DD})$_1$  and  (\ref{E3DDD}) are valid when  the terms in $\dot{(\nabla\mathbf{J}^{(q)})}^{s}$ and  $(\nabla\times \dot{\mathbf{J}}^{(q)})$ are null.
Inserting  equations  (\ref{E3DD})  and (\ref{E3DDD}) in  (\ref{E3Ba}),  we obtain   the 
 following  generalized heat equation valid  in a the steady state
\begin{equation} \label{E10}  
 \mathbf{J}^{(q)} =  - \lambda_1  \nabla T   + l_1^2 \left[ \nabla \left( \nabla\cdot \mathbf{J}^{(q)}\right) +\Delta\mathbf{J}^{(q)} \right]  
- l_2^2 \nabla\times\left( \nabla\times\mathbf{J}^{(q)}\right), 
\end{equation}
where  
\begin{equation}\label{cost}
 l_1^2 = \frac{1}{2}\lambda_1 T^2 \lambda_2^{s} (\beta_1^{s})^2=\frac{1}{2}\lambda_1 T^2  \beta_1^{s},\quad
 l_2^2 = \lambda_1 T^2\lambda_2^{(a)} (\beta_1^{(a)})^2= \lambda_1 T^2 \beta_1^{(a)}, 
\end{equation} 
 are positive.
 \par  
 Equation  (\ref{E10}) is a heat  equation valid in the conditions 
 (\ref{E3Bccc}).
\section{A posteriori identification  to derive   a more descriptive  heat equation seen as asymptotic expansion}\label{sec.asymp}
In this section, we provide an alternative point of view on the a posteriori identification of the internal variables (carried out in the preceding section). The identification can be obtained by approximately solving evolution equations \eqref{eq.evo} while keeping the zero-th and first orders.

To obtain an approximate solution of the evolution equations \eqref{eq.evo}, we need rewrite them to a nondimensional form, using $\mathbf{x} = \bar{x} \hat{\mathbf{x}}$, where $\bar{x}$ stands for a typical length at which spatial variations take place and $\hat{\mathbf{x}}$ is the non-dimensional position,  $t = \bar{t} \hat{t}$ for a typical time   $\bar{t}$ and a nondimensional time  $\hat{t}$ , $T=\bar{T}\hat{T}$ for temperature, $\nabla =   \bar{\nabla}\hat{\nabla} =\frac{1}{\bar{x}} \hat{\nabla}$ for the gradient operator. 
Similarly, we use
$$\mathbf{J}^{(q)} =  \bar{\mathbf{J}}^{(q)} \hat{\mathbf{J}}^{(q)} = \lambda_1 \bar{T}/\bar{x} \hat{\mathbf{J}}^{(q)}, \;\;\text{with}\;\;\hat{\mathbf{J}}^{(q)}= \hat{T}^2 \hat{\nabla}\frac{1}{\hat{T}}, \quad 
  \mathbf{Q}^s =   \bar{\mathbf{Q}^s } \hat{\mathbf{Q}^s} = \frac{\lambda_2^s \beta_1^s \bar{\mathbf{J}}^{(q)} }{\bar{x}} \hat{\mathbf{Q}}^s=\lambda_2^s \beta_1^s \lambda_1 \bar{T}/\bar{x}^2 \hat{\mathbf{Q}}^s ,$$
\begin{equation}\label{nondimensionalquantities}
\mathbf{Q}^{(a)} =   \bar{\mathbf{Q}}^{(a)} \hat{\mathbf{Q}}^{(a)}  =   \frac{\lambda_2^{(a)} \beta_1^{(a)}  \bar{\mathbf{J}}^{(q)} }{\bar{x}} \hat{\mathbf{Q}}^{(a)}  = \lambda_2^{(a)} \beta_1^{(a)}  \lambda_1 \bar{T}/\bar{x}^2\hat{\mathbf{Q}}^{(a)}
 \end{equation}
 for  the heat flux  $\mathbf{J}^{(q)}$ (see (\ref{E3Ba})),   for $\mathbf{Q}^s$ (see (\ref{E3Bb})) and  $\mathbf{Q}^{(a)}$
  (see (\ref{E3Bc})), respectively.
 Hats denote the nondimensional quantities while bars identify the typical scales. 
Tensor fields $\mathbf{Q}^s$ and $\mathbf{Q}^{(a)}$, which result from their evolution equations, can be then formally expanded as the zero-th order contributions, first-order contributions, etc, as in the Chapman-Enskog expansion \cite{DGM},
\begin{subequations} \label{eq.Q.asymp}
\begin{align}
    \label{eq.Q.asymp.a}    \hat{\mathbf{Q}}^s =& \left(\hat{\mathbf{Q}}^s\right)^{(0)} + \frac{\tau_2}{\bar{t}} \left(\hat{\mathbf{Q}}^s\right)^{(1)} + \left(\frac{\tau_2}{\bar{t}}\right)^2\left(\hat{\mathbf{Q}}^s\right)^{(2)} + \mathcal{O}\left(\frac{\tau_2}{\bar{t}}\right)^3\\
    \label{eq.Q.asymp.b}\hat{\mathbf{Q}}^{(a)} =& \left(\hat{\mathbf{Q}}^{(a)}\right)^{(0)} + \frac{\tau_3}{\bar{t}} \left(\hat{\mathbf{Q}}^{(a)}\right)^{(1)} + \left(\frac{\tau_3}{\bar{t}}\right)^2 \left(\hat{\mathbf{Q}}^{(a)}\right)^{(2)} + \mathcal{O}\left(\frac{\tau_3}{\bar{t}}\right)^3
\end{align}
\end{subequations}
for $\tau_2/\bar{t}\ll 1$ and $\tau_3/\bar{t}\ll 1$.
\par  Also,  this approach  falls within the perturbation theory, which  is a  part of asymptotic approximation (see \cite{AG}). Thus, we look for the solution of evolution equations \eqref{eq.evo} as  asymptotic series of  powers of a small parameter, say 
$\varepsilon,$ ($\varepsilon=\frac{\tau_2}{\bar{t}}$ in  (\ref{eq.Q.asymp.a}) and  
$\varepsilon = \frac{\tau_3}{\bar{t}} $ in  (\ref{eq.Q.asymp.b})),
namely with respect to the asymptotic sequence $\left\lbrace 1,\varepsilon, \varepsilon^2,..., \right\rbrace $,   as $\varepsilon \rightarrow 0$.
\par The tensor fields can be expanded simultaneously because their evolution equations are not coupled with each other, so only the ratios of their time scales and the typical time scale of $\mathbf{J}^{(q)}$ are important. The evolution for $\mathbf{J}^{(q)}$ is assumed to proceed at a much slower time scale. 

The evolution equations   (\ref{E3Bb}) and  (\ref{E3Bc})  of $\dot{\mathbf{Q}}^s$ and 
$\dot{\mathbf{Q}}^{(a)}$,   take the  following  nondimensional form   
\begin{equation}\label{nondimesional1}
\frac{\tau_2}{\bar{t}} \frac{\partial}{\partial \hat{t}} \hat{\mathbf{Q}}^s= 
-\hat{\mathbf{Q}}^s - \left(\hat{\nabla}\hat{\mathbf{J}}^{(q)}\right)^s,
\end{equation}
\begin{equation}\label{nondimesional2}
\frac{\tau_3}{\bar{t}} \frac{\partial}{\partial \hat{t}} \hat{\mathbf{Q}}^{(a)}= 
-\hat{\mathbf{Q}}^{(a)}  + \hat{\nabla}\times\hat{\mathbf{J}}^{(q)},
\end{equation}

and they  can be then expanded   by plugging relations \eqref{eq.Q.asymp}   and  the   temporal terivatives with respect to  $\hat{t}$  of  \eqref{eq.Q.asymp}   into them, as
\begin{subequations}\label{eq.reduced.expanded}
\begin{align}
\frac{\tau_2}{\bar{t}} \frac{\partial}{\partial \hat{t}}  \left(\hat{\mathbf{Q}}^s\right)^{(0)} + \mathcal{O}\left(\frac{\tau_2}{\bar{t}}\right)^2 =&
-\left(\hat{\mathbf{Q}}^s\right)^{(0)} - \frac{\tau_2}{\bar{t}}\left(\hat{\mathbf{Q}}^s\right)^{(1)} - \left(\hat{\nabla}\hat{\mathbf{J}}^{(q)}\right)^s,\\
\frac{\tau_3}{\bar{t}} \frac{\partial}{\partial \hat{t}}  \left(\hat{\mathbf{Q}}^{(a)}\right)^{(0)} + \mathcal{O}\left(\frac{\tau_3}{\bar{t}}\right)^2 =&
-\left(\hat{\mathbf{Q}}^{(a)}\right)^{(0)} - \frac{\tau_3}{\bar{t}}\left(\hat{\mathbf{Q}}^{(a)}\right)^{(1)} + \hat{\nabla}\times\hat{\mathbf{J}}^{(q)},
\end{align}
where we assume that the time scales of $\mathbf{Q}^{s}$ and $\mathbf{Q}^{(a)}$ are much smaller than the time scale of $\mathbf{J}^{(q)}$. 
\end{subequations}
In other words, both $\mathbf{Q}^{s}$ and $\mathbf{Q}^{(a)}$ evolve on their fast time scales exponentially to their respective quasi-equilibria
\begin{equation}\label{zerorder}
\left(\hat{\mathbf{Q}}^s\right)^{(0)} = - \left(\hat{\nabla}\hat{\mathbf{J}}^{(q)}\right)^s
\quad\mbox{and}\quad
\left(\hat{\mathbf{Q}}^{(a)}\right)^{(0)} = \hat{\nabla}\times \hat{\mathbf{J}}^{(q)},
\end{equation}
which are the zero-th order solutions of the evolution equations for $\mathbf{Q}^{s}$ and $\mathbf{Q}^{(a)}$, seen also in Equations \eqref{eq.Q0}.
When these relations are plugged back into \eqref{eq.reduced.expanded}, the first-order terms give that
\begin{equation}\label{firstorder}
- \frac{\partial}{\partial \hat{t}}\left(\hat{\nabla}\hat{\mathbf{J}}^{(q)}\right)^s = -\left(\hat{\mathbf{Q}}^s\right)^{(1)}
\quad\mbox{and}\quad
 \frac{\partial}{\partial \hat{t}}\left(\hat{\nabla}\times \hat{\mathbf{J}}^{(q)}\right) = -\left(\hat{\mathbf{Q}}^{(a)}\right)^{(1)}.
\end{equation}
To derive  the constitutive relations for the tensor fields  
 $\hat{\mathbf{Q}}^s$  and     $\hat{\mathbf{Q}}^{(a)}$  up to the first-order, firstly, we have to insert     the zero-th order  and first-order   contributions    (\ref{zerorder})  and (\ref{firstorder}) in (\ref{eq.Q.asymp.a})  and    (\ref{eq.Q.asymp.b}), obtaining   
 \begin{subequations}
\begin{equation} \label{firstresult}
    \hat{\mathbf{Q}}^s =  - \left(\hat{\nabla}\hat{\mathbf{J}}^{(q)}\right)^s +  \frac{\tau_2}{\bar{t}}  \frac{\partial}{\partial\hat{t}}\left(\hat{\nabla}\hat{\mathbf{J}}^{(q)}\right)^s +
    \mathcal{O}\left(\frac{\tau_2}{\bar{t}}\right)^2,
    \end{equation}
    \begin{equation}\label{secondresult}
    \hat{\mathbf{Q}}^{(a)} = \hat{\nabla}\times \hat{\mathbf{J}}^{(q)}+ \frac{\tau_3}{\bar{t}}  \frac{\partial}{\partial\hat{t}}\left(\hat{\nabla}\times \hat{\mathbf{J}}^{(q)}\right)+
    \mathcal{O}\left(\frac{\tau_3}{\bar{t}}\right)^2,
\end{equation}
 \end{subequations}
 then   we  obtain the evolution equations for   the  wanted fields (\ref{E3C}) and  (\ref{E3D}) by the transformations from the nondimensional fields    $\hat{\mathbf{Q}}^s$  and     $\hat{\mathbf{Q}}^{(a)}$  to the fields    ${\mathbf{Q}}^s$  and     ${\mathbf{Q}}^{(a)},$ i.e. multiplying  (\ref{firstresult}) and (\ref{secondresult})  by   $\bar{\mathbf{Q}}^s$ and  by  $\bar{\mathbf{Q}}^{(a)}$, respectively. 
  Thus, we have   
 \begin{subequations}\label{B}
\begin{align}
    \mathbf{Q}^s =& -\lambda_2^s \beta_1^s \left(\nabla \mathbf{J}^{(q)}\right)^s  + \tau_2 \lambda_2^s \beta_1^s \left(\nabla \dot{\mathbf{J}}^{(q)}\right)^s + \mathcal{O}\left(\frac{\tau_2}{\bar{t}}\right)^2,\\
    \mathbf{Q}^{(a)} =& \lambda_2^{(a)} \beta_1^{(a)} \nabla \times \mathbf{J}^{(q)} - \tau_3 \lambda_2^{(a)} \beta_1^{(a)} \nabla\times \dot{\mathbf{J}}^{(q)} + \mathcal{O}\left(\frac{\tau_3}{\bar{t}}\right)^2. 
\end{align}
\end{subequations}
\par From the perspective of the perturbation theory,  we assume $\left(\hat{\mathbf{Q}}^s\right)^{(0)} $  and  $\left(\hat{\mathbf{Q}}^{(a)}\right)^{(0)}$ as the initial steady states of the perturbed   wanted  fields  $\hat{\mathbf{Q}}^s$ and   $\hat{\mathbf{Q}}^{(a)}$ (see (\ref{eq.Q.asymp})), we calculate the partial time derivatives of the asymptotic  expansions  of these fields. 
 Upon inserting the asymptotic expansions (\ref{eq.Q.asymp}) and their partial time derivatives into the constitutive relations  (\ref{nondimesional1}) and  (\ref{nondimesional2})  in the non-dimensional form, \textit{ matching  the obtained  series}, 
the first-order and second-order constitute relations (\ref{zerorder})  and (\ref{firstorder}) follow. Thus, 
by introducing these results  in the asymptotic  expansions (\ref{eq.Q.asymp}), we derive the evolution equations for the wanted fields, applying the same procedure as above. This asymptotic approach  gives an approximate solution of the equations and also provides a method to \textit{determine} the internal variables, and it seems to be used for the first time for the identification of internal variables.
Constitutive relations   (\ref{B})  can be now plugged into the evolution equation    (\ref{E3Ba})  for $\mathbf{J}^q$, which leads to (see also (\ref{gradq})) 
\begin{multline} \label{E10b}  
    \tau_1 \dot {\mathbf{J}}^{(q)} + \mathbf{J}^{(q)} =  - \lambda_1  \nabla T   + l_1^2 \left[ \nabla \left( \nabla\cdot \mathbf{J}^{(q)}\right) +\Delta\mathbf{J}^{(q)} \right]  - h_1^2 \left[ \nabla \left( \nabla\cdot \dot{\mathbf{J}}^{(q)}\right) +\Delta\dot{\mathbf{J}}^{(q)}\right]\\
- l_2^2 \nabla\times\left( \nabla\times\mathbf{J}^{(q)}\right)  + h_2^2 \nabla\times\left( \nabla\times\dot{\mathbf{J}}^{(q)}\right) , 
\end{multline}
where  $$\quad l_1^2 = \frac{1}{2}\lambda_1 T^2 \lambda_2^{s} (\beta_1^{s})^2,\quad h_1^2 = \frac{1}{2}\rho T^2 \lambda_1 \lambda_2^{s} \beta_1^{s}\alpha_2^{s} = \frac{1}{2} T^2 \lambda_1  \beta_1^{s}\tau_2,$$
\begin{equation}\label{costb}
 l_2^2 = \lambda_1 T^2\lambda_2^{(a)} (\beta_1^{(a)})^2, 
 \quad  h_2^2 = \frac{1}{2}\rho\lambda_1 T^2\lambda_2^{(a)} \beta_1^{(a)}\alpha_2^{(a)}= \frac{1}{2}T^2\lambda_1  \beta_1^{(a)}\tau_3
 \end{equation} 
 are positive.
 \par To  obtain  Equation (\ref{E10b})  
  we have taken into consideration that  
 $$\nabla\cdot\left[  \left( \nabla\mathbf{J}^{(q)}\right)^{s}\right]  =  \frac{1}{2}  \nabla \left( \nabla\cdot \mathbf{J}^{(q)}\right)   +  \frac{1}{2}\Delta \mathbf{J}^{(q)}, \quad  \nabla\cdot\left[  \left( \nabla\dot{\mathbf{J}}^{(q)}\right)^{s}\right]  =  \frac{1}{2}  \nabla \left( \nabla\cdot\dot{\mathbf{J}}^{(q)}\right)   +  \frac{1}{2}\Delta \dot{\mathbf{J}}^{(q)},$$ 
 $$\quad \text{with} 
 \quad \left[  \left( \nabla\dot{\mathbf{J}}^{(q)}\right)^{s}\right]_{ij} = 
 \frac{1}{2}\left[ \left( \dot{\mathbf{J}}^{(q)}\right)_{i,j} +  \left( \dot{\mathbf{J}}^{(q)}\right)_{j,i}\right] \quad \text{and} 
$$
$$
 \quad\left[ \nabla\cdot\left(\nabla\dot{\mathbf{J}}^{(q)}\right)^{s}\right]_{i}=  
  \sum_{j=1}^3\frac{\partial}{\partial x_j}\frac{1}{2}\left[\left(\dot{\mathbf{J}}^{(q)}\right)_{i,j} +  
  \left(\dot{\mathbf{J}}^{(q)}\right)_{j,i}\right] = $$
  \begin{equation}\label{gradq} 
  \frac{1}{2}\left[
  \dot{\mathbf{J}}^{(q)}_{i,jj} \; + \; 
\dot{\mathbf{J}}^{(q)}_{j,ji}\right] = 
\frac{1}{2}  \left[\Delta \dot{\mathbf{J}}^{(q)} +
  \nabla \left( \nabla\cdot\dot{\mathbf{J}}^{(q)}\right) \right]_i. 
  \end{equation}
  \par Furthermore,  (\ref{E10b}), with  (\ref{costb}), can take the following form  
  \begin{equation}\label{HE1}
  \tau_1 \dot{\mathbf{J}}^{(q)}= - \mathbf{J}^{(q)} -\lambda_1 \nabla T
    +\gamma^s\nabla(\nabla\cdot \mathbf{J}^{(q)})
    +\gamma^{(a)}\Delta \mathbf{J}^{(q)}
    +\delta^s\nabla(\nabla\cdot \dot{\mathbf{J}}^{(q)})
    +\delta^{(a)}\Delta \dot{\mathbf{J}}^{(q)}, 
 \end{equation}
 with
\begin{equation} \label{HE2}
    \delta^s = \lambda_1 T^2\left(\frac{1}{2}\left(\beta_1 ^{s}\right)^2\lambda_2^{s}\tau_2+(\beta_1^{(a)})^2\lambda_2^{(a)}\tau_3\right)
    \quad\mbox{and}\quad
    \delta^{(a)} = \lambda_1 T^2\left(\frac{1}{2}\left(\beta_1 ^{s}\right)^2\lambda_2^{s}\tau_2-(\beta_1^{(a)})^2\lambda_2^{(a)}\tau_3\right).
\end{equation}
 Of course, when the relaxation times $\tau_2$ and $\tau_3$ are much smaller than the typical time $\bar{t}$, equations (\ref{HE1})  and (\ref{HE2}) simplify to the zero-th order evolution equation for $\mathbf{J}^{(q)}$ \eqref{eq.reduced}. \\
In (\ref{E10b})  the terms in $h_1$ and $h_2$  may be of interest in processes where the time derivative   of $ \mathbf{J}^{(q)}$ is sufficiently high; otherwise, in slow phenomena or in steady states the terms in $h_1$ and $h_2$ disappear and one recovers the enlarged version of the Guyer-Krumhansl equation with   shear and rotational phonon viscosities. 
\par  
 Equation  (\ref{E10b}) is a generalized Guyer-Krumhansl equation more general than the heat transport equation derived in  \cite{1}.  
In Equation (\ref{E10b}),  the coefficients $h_1^2$  and  $h_2^2 $  are proportional to some relaxation time  ($\tau_2$ and $\tau_3$, respectively) times  the square of a mean free path of phonons:  $h_1^2$  and $h_2^2 $ are multiplied by terms describing   particular motions of phonons, see \cite{24},  where the phonons are treated as a Bose gas within a multiscale thermodynamics  and have vortical motions.
 \par  From another perspective, the  thermodynamic formulation given here to describe viscous motions of phonons, where we have used as internal variable the antisymmetric part of  $\mathbf{Q}$,  is analogous  to the formulation  done in hydrodynamics of "micropolar fluids",  composed of elongated molecules (not of spherical form), having rotational degrees of freedom,   in addition to vibrational ones,  around their equilibrium position. This rotational motion  of the molecules,  giving rise to a  rotational viscosity,   can  interact with their vortical motions. In  the same way the phonons can rotate, in addition to their vibration, around the their equilibrium position, as it  can happen  in  ferroelectric  or/and   ferromagnetic solids having  different  molecular species  constituting their crystal lattice,  and the phonon  rotation  can interact with the phonon vortices (for a phenomenological description of these crystals with internal structure, see \cite{Mau77b},  \cite{Mau76a}, \cite{Mau76b}). Moreover, it is possible to have viscous motions  due to  particular material forces created by  the defects in solid media \cite{Mau95}.
\par Equation (\ref{E10}), when the coefficients  $h_1^2 $  and  $h_2^2 $ can be disregarded,  coincides with the heat transport equation (\ref{E102L})    obtained in \cite{1},   where there are not present 
the two  additional   terms  \\ $h_1^2 \left[ \nabla \left( \nabla\cdot \dot{\mathbf{J}^{(q)}}\right) +\Delta\dot{\mathbf{J}}^{(q)}\right]$ and   $h_2^2 \nabla\times(\nabla\times\dot{\mathbf{J}}^{(q)})$,  describing  viscous  motions  and  special vortical motions of phonons, respectively.
\par Equation (\ref{E10}), in the case where    the coefficients  $h_1^2$,   $h_2^2 $ and $l_2^2 $  can be neglected, reduces to Guyer-Krumhansl  heat equation
In the case  when  the coefficients  $l_1^2,  $  $h_1^2,  $  $l_2^2 $  and  $h_2^2 $ 
 can be disregarded, Equation (\ref{E10G}) gives the Maxwell-Vernotte-Cattaneo model (\ref{E10M})  and  
 in the case where the coefficients $\tau,$  $l_1^2,  $  $h_1^2,  $  $l_2^2 $  and  $h_2^2 $ 
 can be neglected, it  reduces  to the Fourier  equation  (\ref{F}).
\section*{Conclusions}
In this paper, we use the Gyarmati  approach within the framework of non-equilibrium thermodynamics with internal variables,  developed in a  previous paper \cite{1}, to describe    particular  motions of phonons,  where  the shear viscosity and also the  rotational viscosity,  in addition to their diffusive and ballistic motions, are taken into consideration. In   \cite{1}, two tensorial internal  variables  were  introduced into the thermodynamic state space to obtain this description,  namely $\mathbf{Q}^{s}$ and   $\mathbf{Q}^{a}$, the symmetric and the antisymmetric part of a second order tensor  $\mathbf{Q}$. 

In this paper, we go further  and  we make the following identifications
$\mathbf{Q}^{s} =  - (\nabla\mathbf{J}^{(q)})^{s}$  and 
 $\mathbf{Q}^{a} =    (\nabla\mathbf{J}^{(q)})^{a}$, or 
 $\mathbf{Q}^{(a)} =   \left(\nabla\mathbf{J}^{(q)}\right)^{(a)}$ for the related axial vector.

By  the identifications   of the two internal tensorial variables   we obtain the  new heat  transport equation  (\ref{E10}), where there are two terms in addition:  the term  $\nabla\times(\nabla\times\dot{\mathbf{J}}^{(q)})$, describing vortical motions of phonons  involving  the time derivation of the heat flux (in analogy to  fluid vortices  involving the acceleration field); and  the other  term    $ h_1^2 \left[ \nabla \left( \nabla\cdot \dot{\mathbf{J}^{(q)}}\right) +\Delta\dot{\mathbf{J}}^{(q)}\right]$  describing particular viscous motions involving   $\dot{\mathbf{J}}^{(q)}$.
Such terms appear for instance in crystals with defects, giving rise   to  material forces   (see  \cite{Mau95}),  generally in   complex solid media, as ferroelectric or/and magnetoelectric crystals,  where  interactions there exist among  the polarizations (or magnetic spins) of the different molecular species constituting the solid  and the crystal lattice of the solid,  and also among  the  polarization gradients (or magnetization gradients of the different  molecules  \cite{Mau77b},  \cite{Mau76a}, \cite{Mau76b}, i.e. this situation  
 may be present when the system is composed of elongated molecules having not only vibration  degrees of freedom but also rotational degrees of freedom around their equilibrium position. Furthermore,  this rotational motion can  interact with the vortical motion of the molecules.
 This could be of interest in the development of new thermal metamaterials, a field of much current interest  \cite{JR2023}. 
 
  Asymptotic reduction provides an alternative point of view  to obtain  the  posteriori identification of the  two internal variables. The latter can be interpreted as the approximate evolution obtained by asymptotically expanding the fast-evolving variables $\mathbf{Q}^s$ and $\mathbf{Q}^{(a)}$ up to the first order around their quasi-equilibria determined by the relatively slowly varying heat flux $\mathbf{J}^{(q)}$. Thus,  when   the relaxation times $\tau_2$ and $\tau_3$ are much smaller than the typical time $\bar{t}$,  using this asymptotic approach,  the  obtained  new generalized heat equation  (\ref{E10b})  is equivalent to equation \eqref{eq.reduced}. 

Equations (\ref{E3B})-(\ref{E3D}) -the central equations of the present paper- are linear; if one goes to a non-linear extension of them, terms in $\mathbf{Q}^a\times \mathbf{J}^{(q)} $ could be added to (\ref{E3B}) and (\ref{E3D}). These terms would be especially relevant in two-dimensional flows, where the vorticity of the heat vortices -related to 
$\mathbf{Q}^a$- has a well-defined direction, perpendicular to the plane of the flow, upwards for counter-clock- wise rotating vortices, downwards in the opposite case. In this case, the terms in 
$\mathbf{Q}^a \times \mathbf{J}^{(q)} $ would describe that the vortices produce a lateral deviation of the heat flux, and that the heat flux produces a lateral deviation of vortices, contributing to a separation between vortices with positive and negative vorticity.  In two-dimensional flows of turbulent superfluids (see for instance \cite{m}, \cite{n}), in which case the vortices are quantized, these effects are well known  \cite{p}, \cite{q},\cite{r}.
\begin{figure}[h]
\centerline{\includegraphics[width=0.3\textwidth]{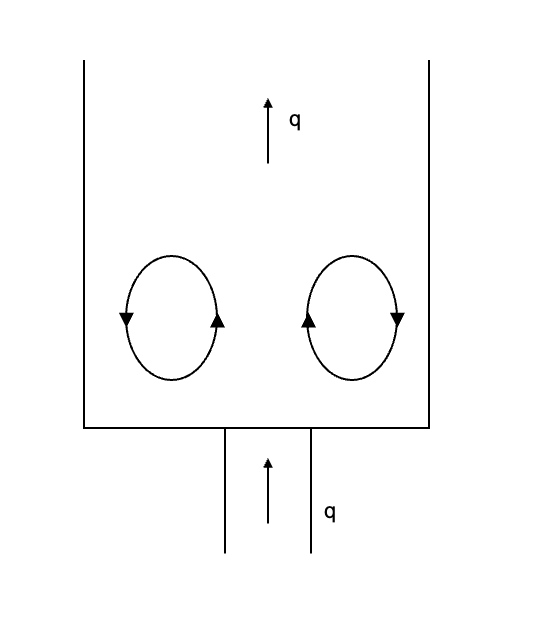}}
\caption{"Possible heat vortices near steep increase  of width"}
\label{fig2}
\end{figure} 
Note that when nonlinear terms are included into   (\ref{E3B})  and (\ref{E3D})    they do not contribute to the entropy production (\ref{E3}), and therefore they may be safely included but do not come directly from the thermodynamic formalism. In  \cite{1}  it was seen that  when  the GENERIC formalism  \cite{24} is used with  additional  hypotheses the Fourier-Crocco equation  may arise, see equation (72) of  \cite{1}, or  equation (50) of  \cite{24},  of the form 
 $$ \;\;\mathbf{J}^{(q)} = - \lambda_1\nabla T - \mathbf{J}^{(q)}\times (\nabla \times \mathbf{J}^{(q)}).$$

Another non-linear extension of equations (\ref{E3B})-(\ref{E3D}) would be considering that the time derivatives of the respective variables contain non-linear convective terms, of the form 
$(\mathbf{J}^{(q)}/\rho c_v T)\cdot \nabla \mathbf{J}^{(q)} $ and so on, where $\mathbf{J}^{(q)}/(\rho c_vT) $ is an average convective speed of phonons, having assumed that    the specific internal energy $u$ is related to the temperature by $u = c_vT$,  with   $c_v$ the heat capacity per unit of mass at constant volume. In this case, interesting effects may arise, as for instance a von-K{\'a}rm{\'a}n street of heat vortices, analogous to the well-known von-K{\'a}rm{\'a}n vortex street in hydrodynamics \cite{24}.

The obtained heat transfer equation   is of special interest in nanosystems in which  $l$  (the mean-free path of the heat carriers,  phonons) is comparable to (or bigger than) $L$ (the  size of the system), 
i.e., for Knudsen number  $\frac{l}{L}\geq1$, where heat vortices may affect the overall heat transfer, as illustrated in Figure \ref{fig2}.

\section*{Acknowledgment}
MP was supported by Czech Science Foundation, project 23-05736S. 
MP is a member of the Ne{\v c}as Centre of Mathematical Modeling.

    \section*{Appendix A}
 In this Appendix we demonstrate that to every antisymmetric tensor, in our case $  \mathbf{Q}^{a},$  it is possible to associate one and only one axial vector   $\mathbf{Q}^{(a)}$  such that 
  \begin{equation}\label{product}  
 Q^{a}_{ij}J_j^{(q)} = (\mathbf{Q}^{(a)}\times\mathbf{J}^{(q)})_i, \quad  \forall \mathbf{J}^{(q)}\;\in\; C, 
  \end{equation}
where $C$ is the thermodynamic state space  and   in matrix formulation  $ \mathbf{Q}^{(a)}$ and   $\mathbf{Q}^{(a)}\times  \mathbf{J}^{(q)}$  have the following form, respectively,\\
  $
Q^{a}_{ij}= 
\begin{pmatrix}
0 & Q^{a}_{12} & Q^{a}_{13}\\
 - Q^{a}_{12}& 0 & Q^{a}_{23}\\
  - Q^{a}_{13} &  -  Q^{a}_{23} & 0
\end{pmatrix},
$ 
 $\quad J_j^{(q)}= (J_1^{(q)}, J_2^{(q)}, J_3^{(q)})^T$, \\
\begin{equation}\label{productA} 
(\mathbf{Q}^{(a)}\times  \mathbf{J}^{(q)})= 
\begin{pmatrix}  
  Q^{(a)}_{2}J_3^{(q)}-Q^{(a)}_{3}J_2^{(q)}\\
Q^{(a)}_{3}J_1^{(q)}-Q^{(a)}_{1}J_3^{(q)}\\
  Q^{(a)}_{1}J_2^{(q)} - Q^{(a)}_{2}J_1^{(q)}
\end{pmatrix}.
\end{equation} 
From expressions   (\ref{productA})  and equation   (\ref{product})  we obtain 
$$  (Q^{a}_{12} + Q^{(a)}_{3})J_2^{(q)} + (Q^{a}_{13} - Q^{(a)}_{2})J_3^{(q)}  = 0, $$
$$ - (Q^{a}_{12} + Q^{(a)}_{3})J_1^{(q)} + (Q^{a}_{23} + Q^{(a)}_{1})J_3^{(q)}  = 0, $$
\begin{equation}\label{productAB}
 - (Q^{a}_{13} -  Q^{(a)}_{2})J_1^{(q)} - (Q^{a}_{23} +  Q^{(a)}_{1})J_3^{(q)}  = 0, 
\end{equation} 
so that  equation (\ref{product}) is equivalent to requiring that each vector $\mathbf{J}^{(q)}$ is a solution of the homogeneous system of equations (\ref{productAB}), i.e.  we have
\begin{equation}\label{Qa}
Q^{(a)}_{1} = - Q^{a}_{23}, \quad   Q^{(a)}_{2} =   Q^{a}_{13},\quad  Q^{(a)}_{3} = - Q^{a}_{12},
\end{equation} 
  or 
\begin{equation}\label{Qab} Q^{(a)}_{i} = - \epsilon_{ijk} Q^{a}_{jk}, 
\end{equation} 
where  $\epsilon_{ijk}$ is the Levi-Civita symbol, equal to zero if   it has two  indeces repeated,  equal to 1  or -1 if the permutation of the indices is   even or odd respect   the fundamental triad 1  2  3. 
    \section*{Appendix B}
  In this  Appendix   we   give the  detailed calculations regarding   the result 
 \begin{equation}\label{nablacdotA}
 \nabla\cdot(\mathbf{Q}\cdot \mathbf{J}^{(q)}) =  \mathbf{J}^{(q)}\cdot(\nabla\cdot Q^T) +  Q  : (\nabla\mathbf{J}^{(q)})^T.
 \end{equation}
 Indeed,  in  components we  have
  \begin{equation}\label{nablacdotB}
  \frac{\partial}{\partial  x_i }(Q_{ij} J_j^{(q)}) =  Q_{ij,i} J_j^{(q)} +   Q_{ij} J_{j,i}^{(q)}=
  (Q_{ji})^T_{,i} J_j^{(q)}  +  Q_{ij} (J_{i,j}^{(q)})^T =  J_j^{(q)} (Q_{ji})^T_{,i} +  Q_{ij} (J_{i,j}^{(q)})^T,
 \end{equation}
 i.e. expression (\ref{nablacdotA}),     beings generally  valid  for a vector $ \mathbf u $  and a second order tensor $\mathbf T$: 
 \begin{equation}\label{nablacdotC} \mathbf T \cdot\mathbf u =  \mathbf u\cdot \mathbf T^T;   \quad   
 T_{ij}u_j = u_j T_{ij}  =  u_j (T_{ji})^T\quad \text{and }\quad 
 \mathbf T :\mathbf T =  T_{ij}T_{ij} = T^T_{ji}T^T_{ji}.  
  \end{equation}

 \section*{Appendix C}
 In the case that we propose as internal variable     $\mathbf{Q}^{a}= (\nabla\mathbf{J}^{(q)})^{a}$,  defined by equation  (\ref{comp})$_2$,  we can  consider the associated  axial vector  $\mathbf{Q}^{(a)} $  as 
\begin{equation}\label{Qac}
\mathbf{Q}^{(a)} =  \frac{1}{2}\nabla\times\mathbf{J}^{(q)}, \quad  
 \text{or in components}\quad {Q}^{(a)}_i =  \frac{1}{2}\epsilon_{ijk}\nabla_j{J}^{(q)}_k = \frac{1}{2}\epsilon_{ijk}{J}^{(q)}_{k,j}. 
\end{equation}
\par Indeed,  taking into account
(\ref{Qac})$_1$,  (\ref{Qa})  and   (\ref{comp})$_2$  we have  \\  
$ \left(\nabla\times\mathbf{J}^{(q)}\right)_1 = 
(\mathbf{J}^{(q)}_{3,2} - \mathbf{J}^{(q)}_{2,3}) =  2Q^{(a)}_{1} = - 2Q^{a}_{23}$, \\
$ \quad\left(\nabla\times\mathbf{J}^{(q)}\right)_2 = 
(\mathbf{J}^{(q)}_{1,3} - \mathbf{J}^{(q)}_{3,1}) =  2Q^{(a)}_{2} =  2Q^{a}_{13}$, \\
$ \quad\left(\nabla\times\mathbf{J}^{(q)}\right)_3 = (\mathbf{J}^{(q)}_{2,1} - \mathbf{J}^{(q)}_{1,2}) =  2Q^{(a)}_{3} = - 2Q^{a}_{12}$. 

 \section*{Appendix D}
In this Appendix we demonstrated the following relation for the divergence of the vectorial product  between  two vectors 
$\mathbf u$  and $\mathbf w$. 
\begin{equation}\label{productACC}
\nabla\cdot(\mathbf u\times \mathbf w) = \mathbf w \cdot\nabla \times \mathbf u  -  \mathbf u\cdot \nabla\times \mathbf w. 
\end{equation}
Indeed,  we can write (\ref{productACC}) in the form
\begin{equation}\label{productACCC}
 \frac{\partial}{\partial x^i} (\epsilon_{ijk} u_jw_k) =  \epsilon_{ijk} u_{j,i} w_k  + \epsilon_{ijk} u_jw_{k,i} =
w_k \epsilon_{kij} u_{j,i}  - u_j \epsilon_{jik}  w_{k,i}.   
\end{equation}
   In the case when  $\mathbf u =  \mathbf Q^{(a)}$ and   $\mathbf w =  \mathbf{J}^{(q)}$ we obtain equation (\ref{exp}).


\end{document}